\documentclass[10pt,journal,compsoc]{IEEEtran}
\IEEEoverridecommandlockouts
\ifCLASSOPTIONcompsoc
  \usepackage[nocompress]{cite}
\else
  \usepackage{cite}
\fi
\usepackage{amsmath,amssymb,amsfonts}
\usepackage{algorithmic}
\usepackage{graphicx}
\usepackage{textcomp}
\usepackage{xcolor}
\usepackage{color}
\def\BibTeX{{\rm B\kern-.05em{\sc i\kern-.025em b}\kern-.08em
    T\kern-.1667em\lower.7ex\hbox{E}\kern-.125emX}}

\usepackage{listings}

\usepackage{pgfplots}
\usepackage{pgfplotstable}
\usetikzlibrary{patterns}
\pgfplotsset{ 
    discard if/.style 2 args={
        x filter/.append={
            \edef\tempa{\thisrow{#1}}
            \edef\tempb{#2}
            \ifx\tempa\tempb
                
            \fi
        }
    },
    discard if not/.style 2 args={
        x filter/.append={
            \edef\tempa{\thisrow{#1}}
            \edef\tempb{#2}
            \ifx\tempa\tempb
            \else
                
            \fi
        }
    },
    compat=1.12
}

\usepackage{ifthen}
\newboolean{showcomments}
\setboolean{showcomments}{true}
\ifthenelse{\boolean{showcomments}}
{ \newcommand{\mynote}[3]{
     \fbox{\bfseries\sffamily\scriptsize#1}
        {\small$\blacktriangleright$\textsf{\emph{\color{#3}{#2}}}$\blacktriangleleft$}}
  }
{ \newcommand{\mynote}[3]{}
  }

\usepackage{hyperref}
\hypersetup{
    colorlinks=true,
    linkcolor=blue,
    filecolor=magenta,
    urlcolor=cyan,
}

\newcommand\copyrighttext{%
  \footnotesize This work has been submitted to the IEEE for possible publication. Copyright may be transferred without notice, after which this version may no longer be accessible.}
\newcommand\copyrightnotice{%
\begin{tikzpicture}[remember picture,overlay]
\node[anchor=south,yshift=10pt] at (current page.south) {\fbox{\parbox{\dimexpr\textwidth-\fboxsep-\fboxrule\relax}{\copyrighttext}}};
\end{tikzpicture}%
}

\begin{document}

\title{PoCL-R: An Open Standard Based Offloading Layer for Heterogeneous Multi-Access Edge Computing with Server Side Scalability}

\author{Jan Solanti, 
Michal Babej,
Julius Ikkala,
Pekka Jääskeläinen
\IEEEcompsocitemizethanks{
\IEEEcompsocthanksitem J. Solanti, M. Babej, J. Ikkala, and P. Jääskeläinen are with Tampere University, Tampere, Finland. E-mail: \{jan.solanti,michal.babej,julius.ikkala,pekka.jaaskelainen\}@tuni.fi}
}

\IEEEtitleabstractindextext{%
\begin{abstract}

We propose a novel computing runtime that exposes remote compute devices via the cross-vendor open heterogeneous computing standard OpenCL and can execute compute tasks on the MEC cluster side across multiple servers in a scalable manner. Intermittent UE connection loss is handled gracefully even if the device's IP address changes on the way.
Network-induced latency is minimized by transferring data and signaling command completions between remote devices in a peer-to-peer fashion directly to the target server with a streamlined TCP-based protocol that yields a command latency of only 60 microseconds on top of network round-trip latency in synthetic benchmarks. The runtime can utilize RDMA to speed up inter-server data transfers by an additional 60\% compared to the TCP-based solution. 
The benefits of the proposed runtime in MEC applications are demonstrated with a smartphone-based augmented reality rendering case study. Measurements show up to 19x improvements to frame rate and 17x improvements to local energy consumption when using the proposed runtime to offload AR rendering from a smartphone.
Scalability to multiple GPU servers in real-world applications is shown in a computational fluid dynamics simulation, which scales with the number of servers at roughly 80\% efficiency which is comparable to an MPI port of the same simulation.
\end{abstract}

}
\maketitle
\IEEEdisplaynontitleabstractindextext
\copyrightnotice

\IEEEraisesectionheading{\section{Introduction}\label{sec:introduction}}

\IEEEPARstart{D}{emand} for various end-user applications to run on battery-powered \textit{User Equipment (UE)} is steadily increasing as such devices have become more portable and commonplace, but at the same time, the applications have become more computationally intensive. One way to solve this dilemma is to offload some computationally intensive parts of the application to a shared compute server cluster located on a nearby edge cloud. This concept is often called \textit{Multi-access Edge Computing (MEC)}~\cite{MEC}.

A key challenge of MEC is needing a way to access both UE-local and MEC-remote heterogeneous compute accelerators  efficiently and seamlessly. The offloading method should be vendor neutral to encompass the wide variety of existing compute accelerators on the market. Furthermore, the parts of the solution that reside in the edge cloud should not be specific to any end-user application, since that complicates server side deployment. Finally, the solution must cope with UE intermittently losing connectivity and roaming between wireless networks.

In this article we propose \textit{PoCL-R},
a computing runtime that acts as a scalable low-latency hardware abstraction layer for distributed heterogeneous compute devices and aims to answer these unique challenges of MEC deployments.

The key benefit of \textit{PoCL-R} in the use of MEC offloading is its API which is solely based on OpenCL~\cite{opencl-api-spec}, an open cross-vendor heterogeneous computing standard which enables accelerator code portability across various UE-local and remote MEC-side devices. \textit{PoCL-R} is targeted for use by application developers directly or as a backend for a number of higher-level APIs like SYCL~\cite{sycl-spec} or CUDA/HIP~\cite{chipstar}. 
Furthermore, the use of OpenCL enables defining the whole application logic in the UE-side host program; the generic PoCL daemon in the MEC server side can accelerate any OpenCL program running in the UE.

The runtime has the following novel features:

\begin{itemize}
  \item Utilization of MEC compute resources with \textit{peer-to-peer (P2P)} communication and synchronization between compute nodes for improved compute scalability.
  \item Capability to support applications with both high performance and low latency demands
  \item Graceful handling of connection loss and restoration on unreliable networks and while roaming.
  \item A minimal (optional) OpenCL API extension that can improve transfer times of dynamic-size buffers dramatically. This is particularly useful for taking advantage of buffers with variable length compressed data.
  \item The first distributed OpenCL runtime that is integrated to a long-maintained widely used open source OpenCL implementation framework \textit{PoCL}~\cite{pocl} and thus is usable and extensible for anyone freely.\footnote{\textbf{Source code available at: \url{http://code.portablecl.org/}}}
\end{itemize}

This article is an extension of a conference publication where we presented an early version of \textit{PoCL-R}~\cite{pocl2021samos}. For this article, we extended the runtime to better solve the challenges posed by more challenging MEC use cases and present additional interesting usage scenarios as follows:
  1) Remote Direct Memory Access (RDMA) support for server-side buffer migrations which bring up to 60\% speedups compared to simple TCP socket communication.
  2) Graceful handling for connection loss between the UE and the MEC to account for roaming and unreliable wireless networks.
  3) Multi-server performance scaling case study that runs the HPC simulation FluidX3D~\cite{Lehmann_FluidX3D_2022} on multiple GPU server nodes.

The new benchmark case demonstrates that \textit{PoCL-R} has potential for offloading unmodified high performance applications that require multiple GPU nodes.

This article is organized as follows:
Section~\ref{section:background} briefly covers the relevant background of MEC and OpenCL.
Section~\ref{section:relatedWork} provides a brief overview of the current available options for offloading at the compute API level.
Section~\ref{section:architecture} describes the \textit{PoCL-R} software architecture on a high level.
Section~\ref{section:runtime} describes the most relevant techniques in the proposed runtime to achieve the low latency while retaining scalability. Section~\ref{section:results} lays out the synthetic latency and throughput benchmark results, and Section~\ref{sec:case-studies} demonstrates real-world performance with two case studies. Finally, Section~\ref{section:conclusions} describes some future improvement plans and presents the conclusions drawn from the benchmarks.

\section{Background}
\label{section:background}

\subsection{Multi-Access Edge Cloud Offloading}

Perhaps the main challenge with MEC offloading is the extra latency caused by transferring data across the access network. Furthermore, if the network is wireless, packet loss, network congestion and varying network conditions frequently hurt the latency even more. As a result, offloading across the network has generally been deemed impractical for many applications until recent developments in wireless networking. 
 
Some prominently advertised features of the current-gen 5G and WiFi6 standards are the improvements they bring to network latency across the board, as well as the quality of service in crowded environments compared to their respective earlier generations. The in-development WiFi7 further aims to help latency-sensitive applications with explicit \textit{Time-Sensitive Networking (TSN)} capabilities~\cite{wifi7}.

In terms of application software support, due to the UE roaming, MEC requires solutions that allow utilizing a diversity of heterogeneous compute resources both locally in the UE, and in encountered MEC servers. The UE software should be possible to be made highly portable in order to retain flexibility of harnessing all available hardware for accelerating the application. Since the needs of each application can vary greatly, the flexibility of the MEC-side service is essential to avoid the need to deploy a custom server side service with each UE application.  

Finally, for optimal performance and resource utilization, the application needs to be able to make intelligent decisions regarding which computations to offload and when. 
In this article, our focus on delivering a device-portable yet general purpose solution for MEC compute execution. Our assumption is that the UE application software has been partitioned to accelerated parts and the invocation logic is performed with another layer such as MRLCO which focuses specifically on use cases where the UE physically moves across environments with different network conditions~\cite{wang2021offloading}. 

\subsection{Open Computing Language}

Open Computing Language (OpenCL) is an open cross-vendor standard for heterogeneous compute~\cite{opencl-api-spec}. Thanks to its native support from multiple hardware vendors, OpenCL is an interesting option for MEC applications and libraries that require portability. 

OpenCL divides the application logic to two parts: The host program running in the host processor and a device-side program consisting of compute kernels.

The OpenCL standard allows specifying kernels in its own dialect of the C language, in a target-specific binary representation or using a target-independent intermediate representation (SPIR-V~\cite{SPIR-V}). There is also an option to invoke so called ``built-in kernels'' which can be used to abstract hardware accelerated functions. 

OpenCL supports online compilation of kernels support, but it's not a required feature; application logic can be defined solely using device-specific binaries or preferably with the portable SPIR-V format. 

OpenCL was originally intended to be utilized for driving local single node computation with the compute distribution is typically delegated to other APIs, typically a Message Passing Interface (MPI). 
In this article, we demonstrate that OpenCL is a good match for distributed MEC offloading
as well. Being able to utilize a single compute API greatly simplifies the UE application logic, since local and remote execution can be done with identical API calls. The capability of OpenCL to describe both the host and the accelerator program in a single software description helps in the goal of having all the program logic in the UE side.

\section{Previous Work}
\label{section:relatedWork}

Offloading and distributing compute workloads has been attempted in the past by multiple projects~\cite{jcl,vocl,vopencl,dopencl} but they have largely faded into obscurity and their implementations are no longer available for comparison nor for further development.
A great number of these projects~\cite{dopencl,clopencl,virtualcl,distcl} mainly focus on HPC clusters and the existing ecosystems and libraries. Their focus is mainly on throughput, and latency is frequently ignored. Our proposed runtime does target high performance compute clusters but also real-time applications and especially aims to combine the two.

Distributing OpenCL operations to networked compute servers as such is not a new idea, however, \textit{PoCL-R} has an explicit focus on latency and client portability while previous solutions have mainly targeted non-real-time computation in data centers.

Out of the related projects we found, the one closest to \textit{PoCL-R} is SnuCL~\cite{snucl_base}. Like \textit{PoCL-R}, it implements the OpenCL standard API and provides facilities for offloading OpenCL commands to remote servers. However, like most pre-existing projects, SnuCL primarily targets HPC clusters and is mainly concerned with throughput. Instead of plain sockets, it uses the MPI framework to handle communication. Some peer-to-peer data transfer functionality exists in SnuCL as well, but the authors report that it has problems with scaling in some tasks, such as a matrix multiplication benchmark that is similar to the one used in this paper. By contrast, \textit{PoCL-R} uses plain TCP sockets with their parameters tuned to reduce latency. The wire representation of commands is kept identical to the in-memory one to avoid a translation step. \textit{PoCL-R} also performs command scheduling independently on the remote servers to the extent possible, whereas SnuCL relies on the client application for this.

There is also a continuation of the SnuCL work, named SNUCL-D~\cite{snucld}, which duplicates the control flow of the entire client application to each remote server. This should improve scalability, but adds a requirement for the host application to be fully replicable on all remote servers, which is additional development work and is obviously not always feasible e.g., with graphical applications targeting mobile devices or if the application accesses local file system resources. 

\textit{PoCL-R} enables performant offloading capabilities to any OpenCL application with no code changes.

In terms of the general concept, rCUDA~\cite{rcuda1} is also similar to \textit{PoCL-R}. At the time of writing, it is the closest related project with the most active development and also supports RDMA via GPUDirect.~\cite{rcuda2} Unfortunately, it is limited in what hardware it can support and which platforms it can be ported to due to being based on the proprietary CUDA API, while \textit{PoCL-R's} key benefit is its cross-vendor portability.

RemoteCL is another OpenCL-based project with a similar approach to \textit{PoCL-R} in that it uses plain TCP sockets for communication. The author of RemoteCL is however not interested in generalizing the library for full OpenCL standard conformance as stated in the project's README file\cite{remotecl}. RemoteCL also seems to be limited to using one remote server at a time, whereas \textit{PoCL-R} can use multiple.

\section{PoCL-Remote Architecture}
\label{section:architecture}

\subsection{Requirements}

\begin{figure}
  \centering
  \includegraphics[width=0.7\columnwidth]{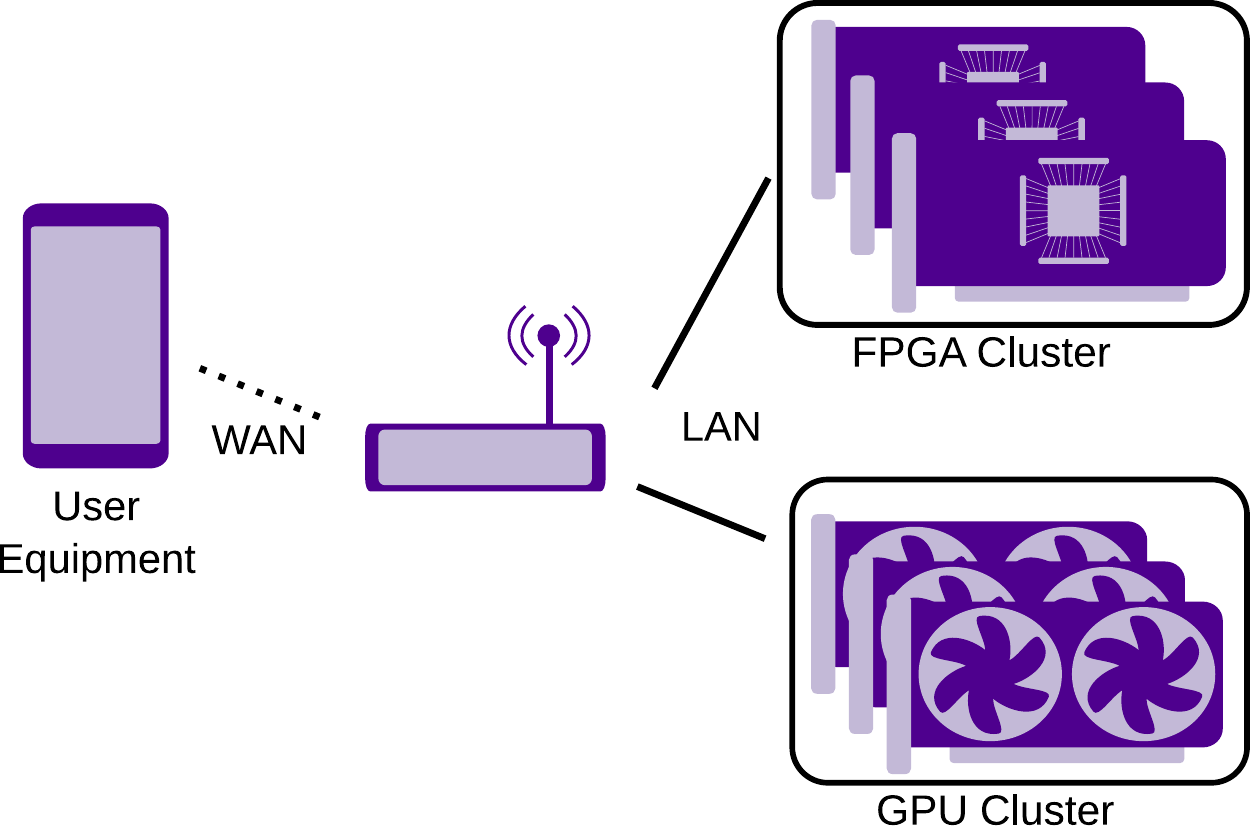}
  \caption{High-level overview of a diverse distributed execution use case where multiple devices with varying types are utilized remotely by a lightweight UE device.}
  \label{fig:pocl-hl}
\end{figure}

Fig.~\ref{fig:pocl-hl} displays a simplified view of a setup where a mobile device runs an application that offloads parts of its computational workload across a low-latency wireless link to one or more clusters comprised of a variety of compute accelerators such as GPUs and FPGAs. This kind of low power UE with a wireless connection to a powerful compute cluster is the main reference scenario for the design of \textit{PoCL-R}.

The primary requirement for \textit{PoCL-R} is to provide an offload layer with minimal end-to-end latency, while the usage of local and remote accelerators is identical. Second, to support UE applications with high performance computing needs, the server-side scalability is essential. Third, 
since the software layer is designed to run on wireless links which can suffer from connection drops, there should be a seamless recovery mechanism presented to the client application.

\subsection{Client-Server Organization}

\begin{figure}
    \centering
    \includegraphics[width=0.7\columnwidth]{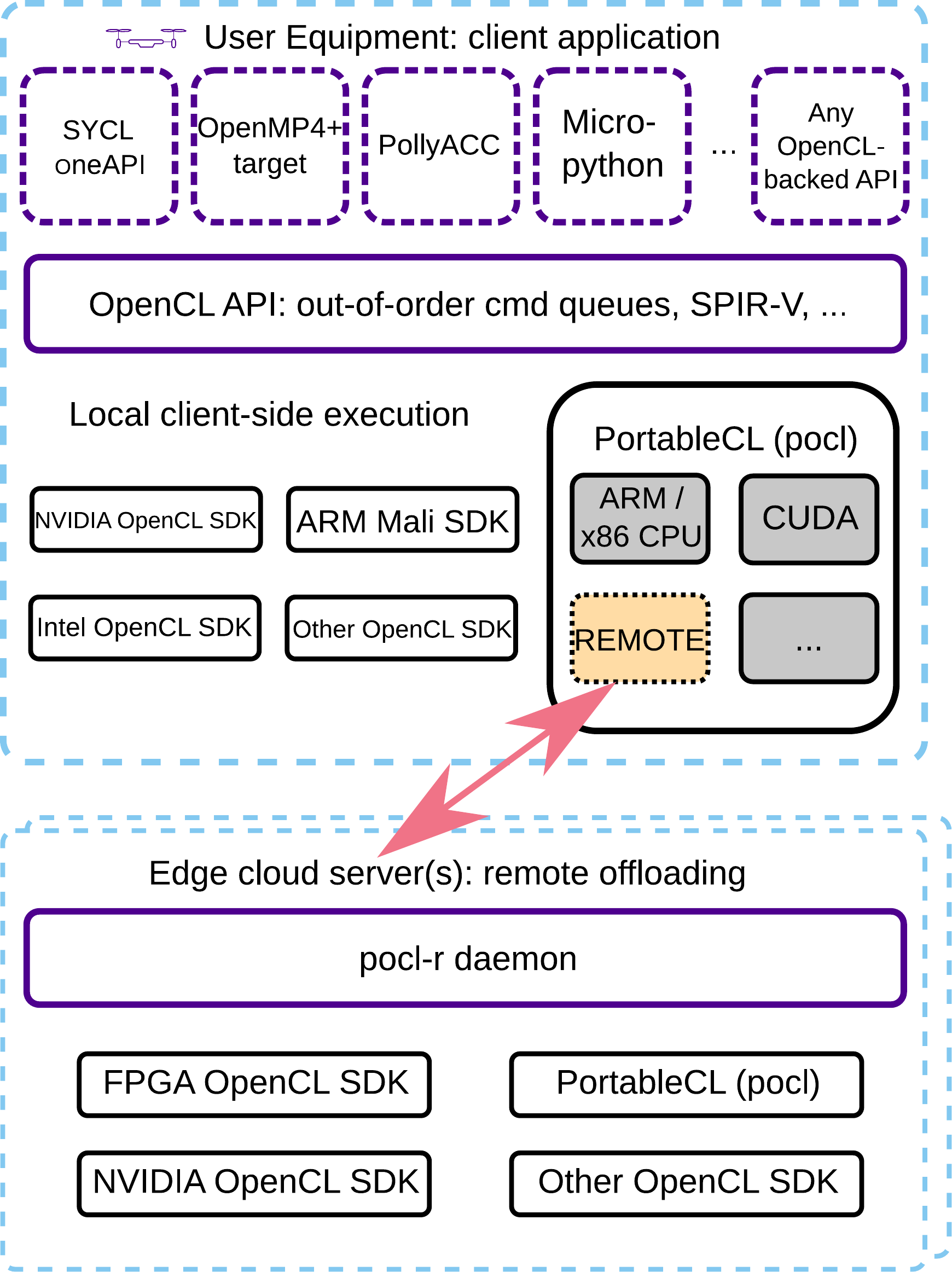}
    \caption{Overview of the software stack for an application using \textit{PoCL-R}. The OpenCL API can be used directly for maximum efficiency, but also as a middleware for improved productivity APIs on top of it. In this stack, \textit{PoCL} is an OpenCL API implementation, a drop-in alternative to the other OpenCL implementations, with the special remote driver interfacing to the remote OpenCL-supported devices with distributed communication.}
    \label{fig:pocl-r-stack}
\end{figure}

At the high level, the \textit{PoCL-R} runtime is built on a client-server architecture, as is common for networked applications. The \textit{client} side of the runtime is implemented as a \textit{remote driver} in the \textit{Portable Computing Language (PoCL)}~\cite{pocl}. PoCL is an open-source implementation of the OpenCL API and has a lot of flexibility for implementing custom backends for diverse hardware and underlying software stacks. The remote driver exposes the compute devices on a given MEC server through the OpenCL platform API, making them appear identically to local devices. 

Fig.~\ref{fig:pocl-r-stack} shows the relations of the various frameworks and drivers on both the client device and compute servers.

The host application using the OpenCL API can use \textit{PoCL-R} as a drop-in implementation without recompilation. When linked against \textit{PoCL-R}, the OpenCL calls are made to the \textit{PoCL-R} \textit{client} driver, which in turn connects to one or multiple \textit{remote} servers, each providing one or more \textit{remote compute devices}. The remote servers can form interconnected \textit{clusters} visible and controlled by \textit{PoCL-R} as \textit{peers} to avoid round-trips back to the client whenever synchronization or data transfers are needed between the remote devices.

The server side is a daemon that runs on the remote servers and receives commands from the client driver, and dispatches them to the OpenCL driver of the server's devices accompanied by proper event dependencies. The OpenCL devices can be controlled via a device-specific proprietary OpenCL driver by the daemon, or through, e.g., the open source drivers provided by \textit{PoCL}.

\begin{figure}
    \centering
    \includegraphics[width=0.7\columnwidth]{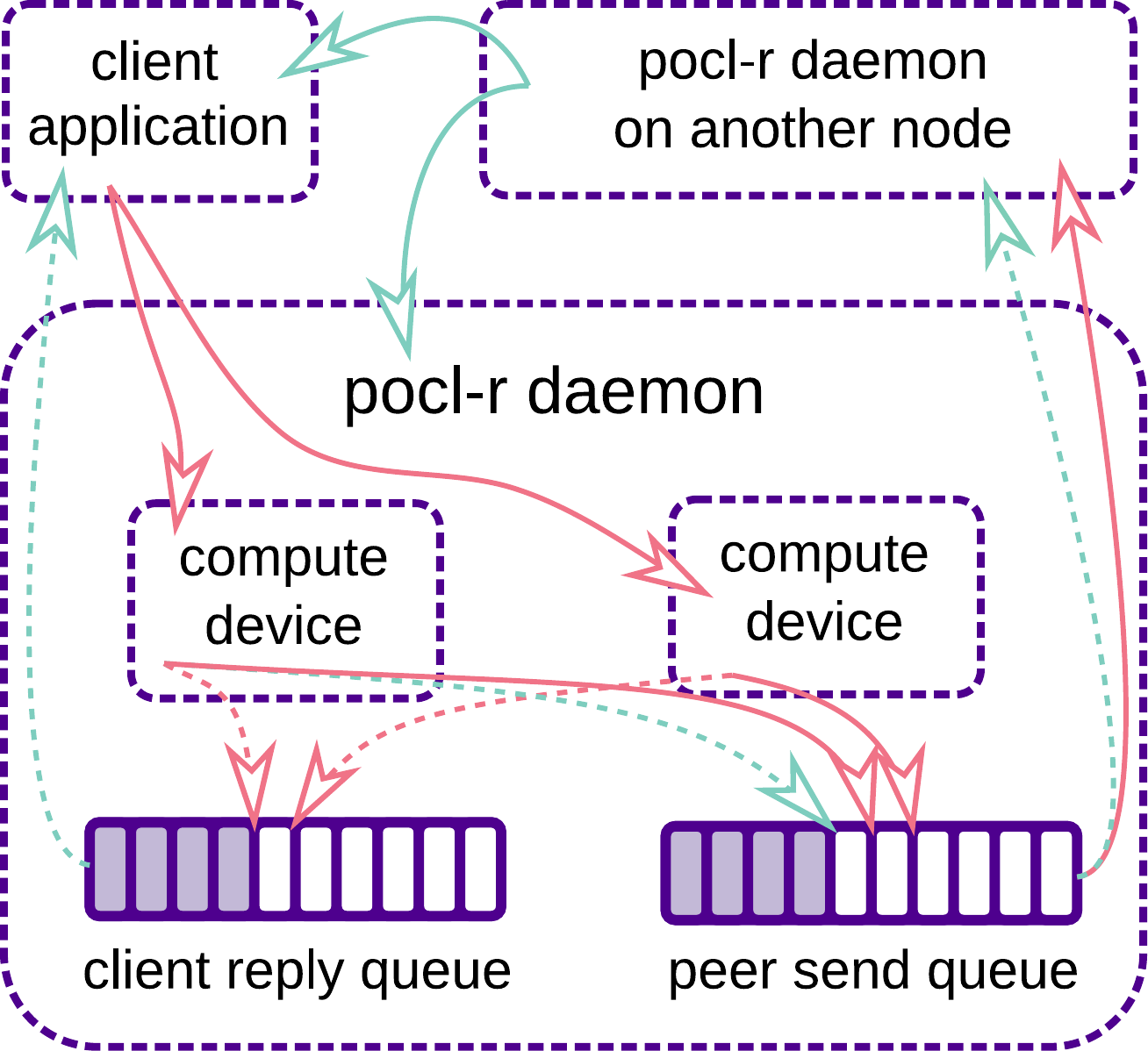}
    \caption{The flow of data and commands from an application to the \textit{PoCL-R} daemon and between remote servers. Red arrows represent OpenCL commands and green arrows indicate command completion notifications. Dashed arrows represent the control flow for a different command whose execution does not require migration to another server, but whose completion is still signaled to other servers.}
    \label{fig:pocl-queues-traffic}
\end{figure}

The daemon is structured around network sockets for the client and peer connections. Each socket has a reader thread and a writer thread. The readers do blocking reads on the socket until they manage to read a new command, which they then dispatch to the underlying OpenCL runtime, store its associated OpenCL event in a queue and signal the corresponding writer thread. The server writer thread iterates through events in the queue and when it finds one that the underlying OpenCL runtime reports as complete, writes its result to the socket representing the host connection. Peer writers have separate queues, but are otherwise similar to the server writer. The server writer adds events that peers need to be notified of to these queues and signals the peer writers. Fig.~\ref{fig:pocl-queues-traffic} illustrates this architecture and the flow of commands and data through it.

\subsection{Handling Connection Loss}

Wireless connections are prone to interference and limited in range, making it necessary to gracefully handle intermittent loss of connection. Handling this well requires designing the networking logic around the assumption that the connection can be lost at any point, even mid-command.

When \textit{PoCL-R} detects that the connection to a remote server is lost, it marks all OpenCL devices associated with that server as unavailable, yielding a "device unavailable" error status from all OpenCL calls that attempt to access such a device after this point. It then attempts to reconnect to the server and once successful, marks the devices as available again and resumes command submissions for that server.

\begin{figure}
    \centering
    \includegraphics[width=0.7\columnwidth]{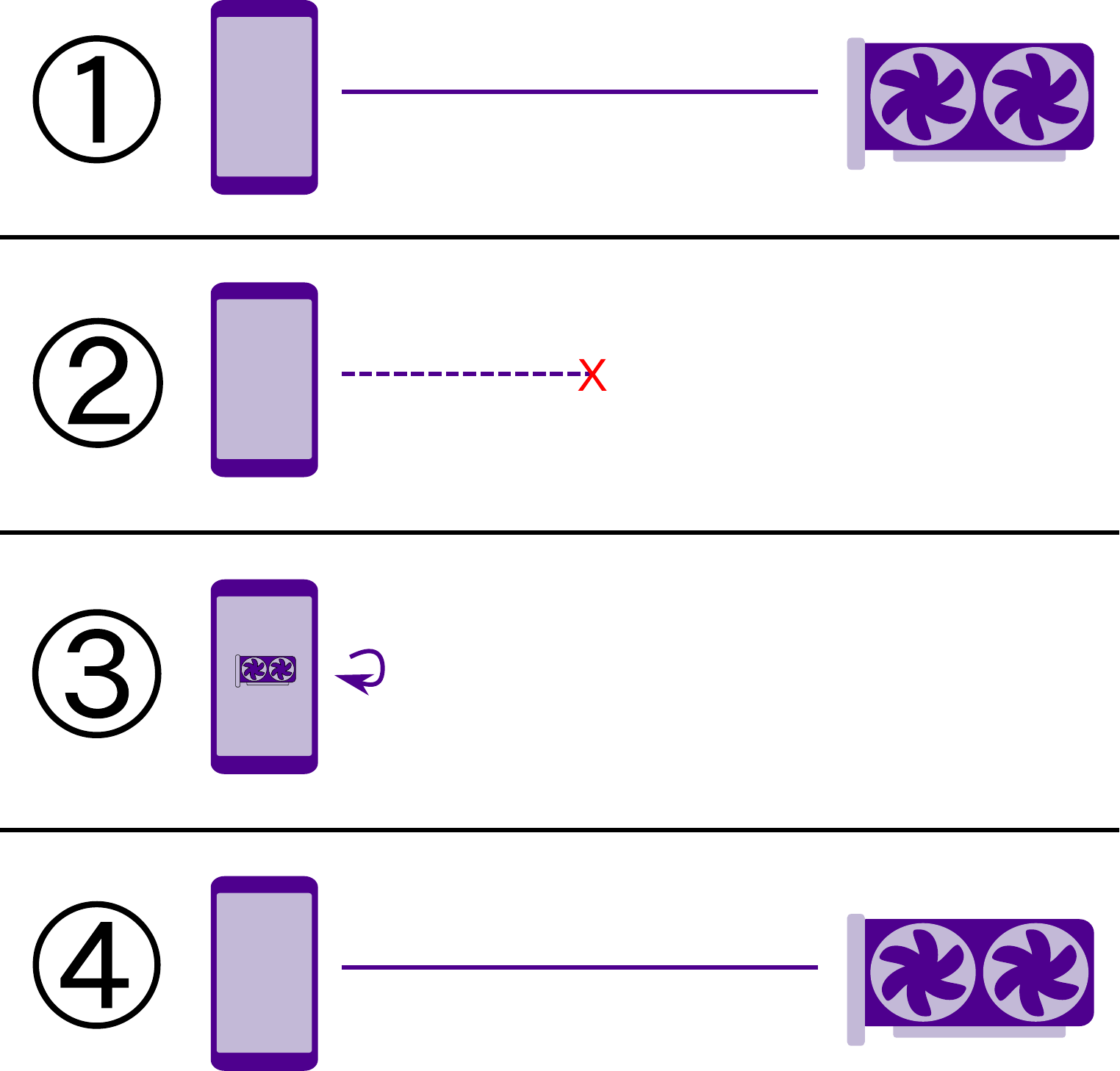}
    \caption{Reconnecting with fallback to local computation. Applications can be designed to primarily use remote devices for computation in order to leverage the latest and most accurate algorithms. When connectivity to the remote compute servers is lost due to network conditions or the UE moving out of the wireless access point's range, applications can fall back to a simpler, less accurate model that is feasible to compute on the mobile SoC. Once remote servers are reachable again, the application can switch back to the full accuracy computations.}
    \label{fig:reconnect-local}
\end{figure}

In the event of a sudden connection loss, it is possible that some commands are partially sent and can't be processed by the server. To account for this, \textit{PoCL-R} keeps a backup of the last few commands it attempted to send and re-sends those before any other commands once a connection is established again. The server simply ignores commands it has already processed.
Higher software layers can make use of the device availability information and manually fall back to performing computations locally, as illustrated in Fig.~\ref{fig:reconnect-local}. Depending on the application, it can make sense to perform local computation in full precision at the cost of greatly increased power consumption and reduced performance. Alternatively, it may be possible to reduce the quality of results by falling back to simpler algorithms, e.g., with lower object detection accuracy. Once the remote devices become available again, computation can be shifted back to those at full precision.

In order to facilitate reliable reconnecting without having to retransmit unchanged data, a 16-byte session ID is introduced. When a client initially connects to a server, it sends a handshake packet containing an all-zeroes session ID. The server generates a random ID and sends it as part of its handshake reply. When reconnecting, the client replaces the all-zeroes ID with the one received from the server it is reconnecting to. The server keeps track of IDs it has handed out and uses them to attach incoming connections to the correct contexts.

\section{Latency and Scalability Optimizations}
\label{section:runtime}

The following subsections describe the latency and scalability enhancing techniques applied in \textit{PoCL-R}.

\subsection{Peer-to-Peer Communication}
\label{sec:p2p}

\begin{figure}
  \centering
  \includegraphics[width=0.7\columnwidth]{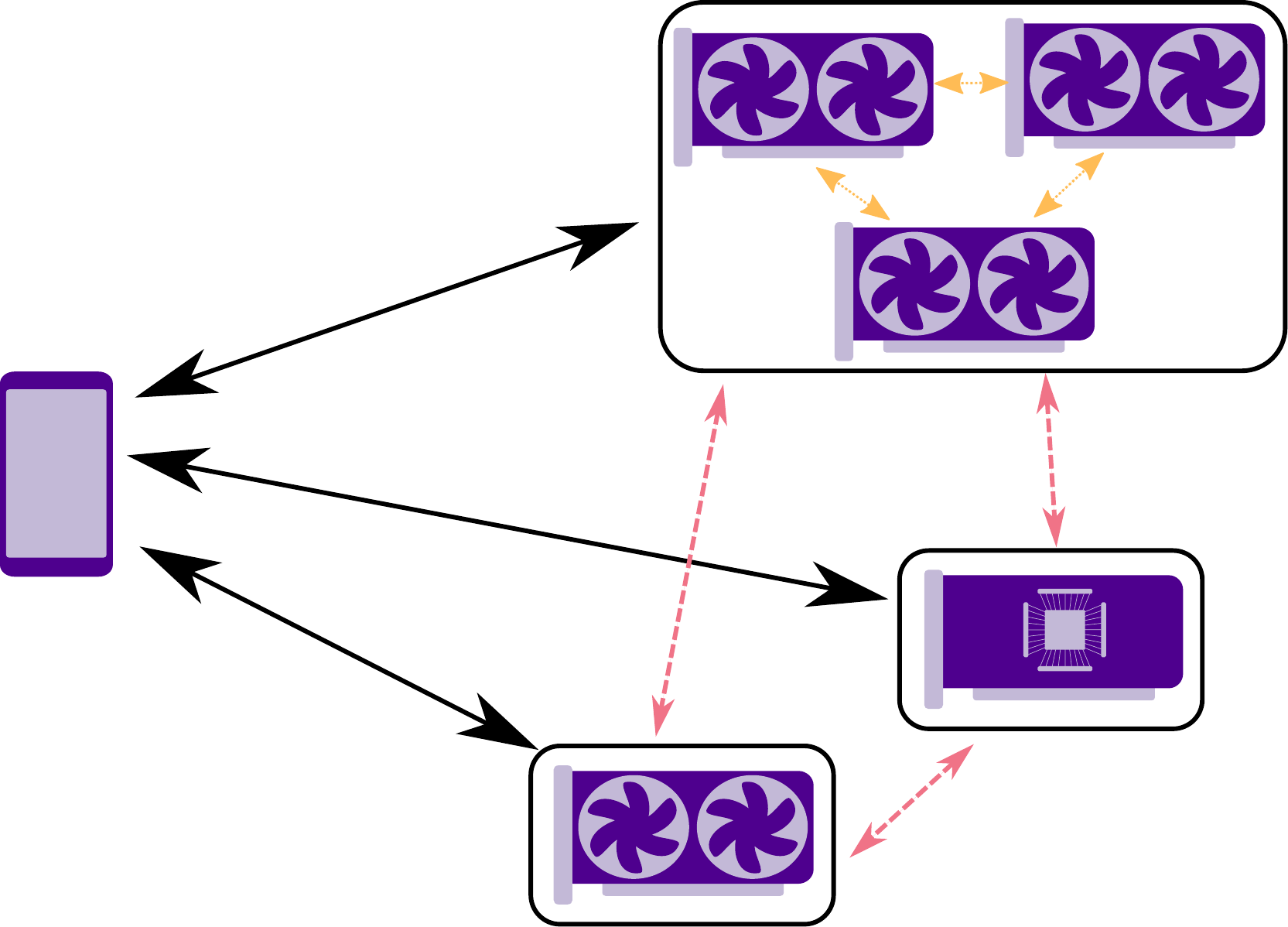}
  \caption{Different paths for buffer transfers supported in \textit{PoCL-R}: P2P network transfers between remote servers in red, direct transfers between devices within a server in yellow. Black arrows denote transfers between the client device and the remote servers.}
  \label{fig:pocl-new}
\end{figure}

Rather than routing all buffer transfers through the client device, \textit{PoCL-R} supports a number of shorter paths when available, as illustrated in Fig.~\ref{fig:pocl-new}. Devices on the same server typically support migrating OpenCL buffers directly between them, which \textit{pocld} will make use of. MEC and data center environments frequently have high-speed interconnects in the form of internal InfiniBand and Ethernet networks, reaching speeds in the order of hundreds of gigabits per second between servers. \textit{PoCL-R} makes use of these connections by transferring buffers and signaling command completions directly between servers in a peer-to-peer (P2P) fashion.

The other part of a typical MEC setup, the UE, on the other hand, is frequently connected wirelessly with limited bandwidth and a nontrivial amount of network congestion or other interference. As a result, network roundtrips to and from the UE ought to be considered relatively expensive and UE bandwidth should be used sparingly. To this end, \textit{PoCL-R} only sends a request to the source server when migrating buffers between servers. The source server then pushes the data directly to the destination, and only the destination server notifies the client of the migration's completion. Compared to the naive solution of downloading the buffer from the source to the client and then uploading it to the destination, this completely eliminates transferring the contents of the potentially very large buffer across the slowest link in the system. One command roundtrip from the client and the wait for a reply from the server is also eliminated.

\subsection{Decentralized Command Scheduling}

When the client application makes an OpenCL API call, the respective command is pushed to the compute server immediately. If the command has a data dependency to prior commands, this is communicated by the application with OpenCL events. The runtime maps the events of the client application to local events of the native context in the remote daemon. Events of commands that are executed on a different server or on the client are mapped to OpenCL user events instead. This way, the task graph defined by the application-specified event dependencies stays intact and can be relied on to perform run time optimizations using the dependency rules presented in~\cite{exploiting_tasks}.

Besides the two connections each server has to the client, servers are directly connected to each other, forming a mesh of peers. These peer-to-peer connections are used for buffer migrations. Additionally, the completion of commands is reported to other servers using these direct connections, as to avoid an unnecessary client roundtrip. Fig.~\ref{fig:pocl-queues-traffic} illustrates the flow of information and control for a command requiring buffer migration and one that does not require migration but sends a completion notification to the other servers:
The client sends a command to the server that has the most up-to-date contents of the needed buffer. This server then directly pushes the buffer to the target server, which in turn sends a command completion to the client and the other servers. If no migration is needed, the client simply sends the command to the target server, which replies to the client and notifies other servers once the command has finished.
Any server that has received a command depending on a command executing on a different server can begin executing such blocked commands immediately when it receives completion notifications for each command being depended on.

\subsection{Dynamic Buffer Content Size Extension}

Device memory in OpenCL contexts is managed in the form of buffers whose size is fixed at creation time. Applications may however work with very varying amounts of data, especially when varying rate compression is involved. In such cases, the buffers allocated need to be sized conservatively, or the application needs to be able to split work into multiple chunks in order to fit a small buffer size. The latter is not necessarily very practical, and as a result a lot of memory can get wasted in preparation for a worst-case scenario. If buffers need to be migrated between devices, this further translates to wasted bandwidth. Network transfers are often orders of magnitude slower than interconnects within a single machine, which greatly emphasizes the overhead of transferring unneeded data.

As a performance improvement for transferring dynamic data sizes, we propose an OpenCL extension named \textit{cl\_pocl\_content\_size}. This extension allows applications to convey to the OpenCL runtime how much of any given buffer is actually used for meaningful data that has to be copied when migrating the buffer. This is done by designating a separate buffer holding exactly one unsigned integer as a ''content size buffer´´. The \textit{PoCL-R} runtime reads this content size buffer and only sends the necessary parts of buffers across the wire when migrating buffers between remotes.

The extension does not alter the way the runtime behaves when an application does not specify a content size buffer.
The extension is discussed more thoroughly in ~\cite{pocl2021samos} and its implementation has been integrated to the PoCL open source repository.

\subsection{RDMA-Based Server-Side Buffer Transfers}

\textit{PoCL-R} can use RDMA to reduce the overhead of buffer migrations between server nodes compared to the original peer-to-peer communication via plain TCP streams.

RDMA is an API with associated networking protocols intended for transferring the contents of memory buffers across a network without CPU involvement and without making extra copies of the data to be sent. In stark contrast to the raw stream of bytes that a TCP socket offers, the RDMA API is designed around the concept of discrete messages managed via \textit{InfiniBand verbs}.\cite{rfc5040rdma, rfc5041rdma} 

The TCP stream communication scheme relies on a union of C structs defining the different commands that can be sent. The in-memory layout is kept consistent by attribute annotations forcing specific packing and alignment rules. However, the command structs vary in size from tens of bytes to multiple kilobytes and unions are always sized based on their largest field, which means every command would take up multiple kilobytes. In memory this is not a big problem, but it becomes a noticeable overhead when transferring commands over the TCP socket.

In order to avoid sending meaningless data alongside commands, the basic communication scheme first sends a standalone size field indicating the size of the specific command being sent, followed by that many bytes of the command union. The unused bytes of the union are left in an undefined state. For commands that require additional data such as buffer contents or OpenCL program build logs, the sizes of these are contained in the command struct and the buffers themselves are sent immediately after the struct.

Having a separate command size at the start of each command means that for each command there will be a minimum of two write calls to the TCP socket, and a minimum of three write calls for a buffer transfer command. When transferring large additional buffers, the socket API sometimes requires splitting the writes up into multiple smaller ones, further increasing the number of system calls.
\begin{figure}
    \centering
    \includegraphics[width=0.7\columnwidth]{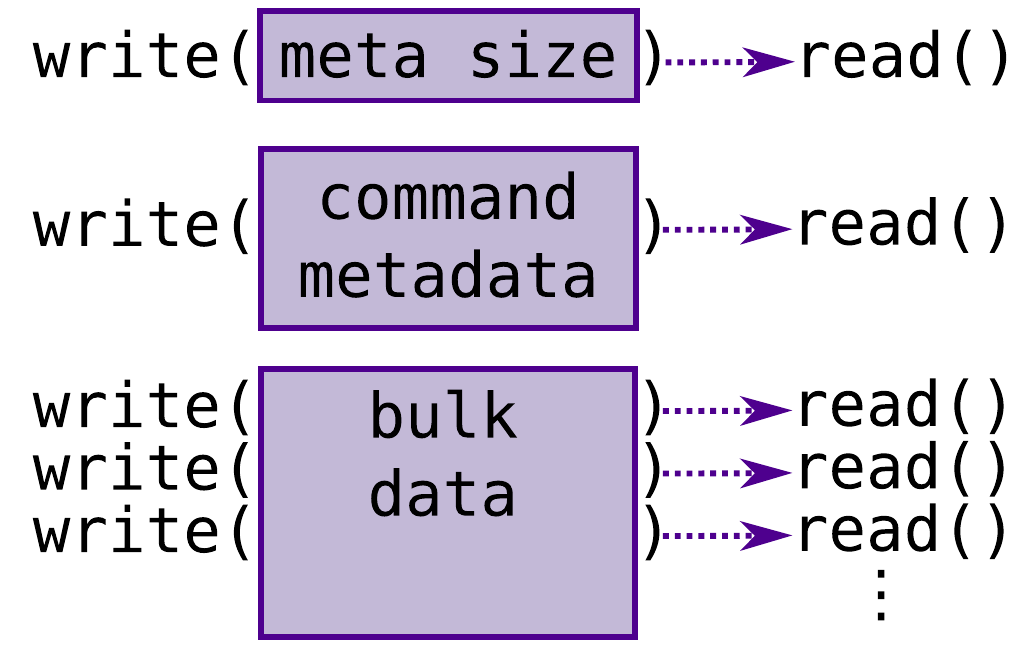}
    \caption{Control flow of transferring a buffer between servers via TCP sockets. Commands are prefixed with a size field in order to avoid wasting bandwidth, as the size of command metadata varies a lot. For commands that require additional data, e.g., buffer contents, the size of this data is part of the command metadata and the data itself is simply written as-is immediately after the command metadata. If any of these parts is larger than the socket's internal buffer, it has to be split into multiple writes/reads, which translates to multiple system calls.}
    \label{fig:tcp_flow}
\end{figure}
\begin{figure}
    \centering
    \includegraphics[width=1.0\columnwidth]{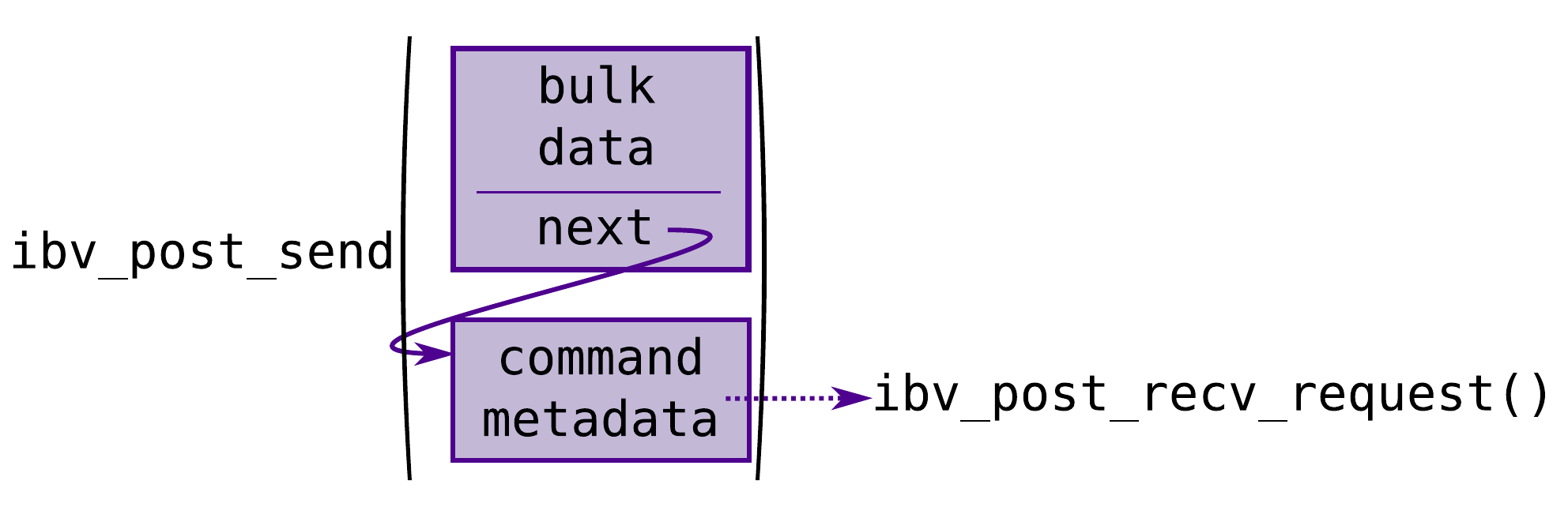}
    \caption{Control flow of transferring a buffer between servers via RDMA. Compared to Fig.~\ref{fig:tcp_flow} no in band size fields are needed since RDMA is message-based and handles message size internally. RDMA messages can also be chained in order to submit multiple work requests with a single function call. The RDMA driver and hardware handle the multiple messages and potential message fragmenting internally without the need for additional syscalls.}
    \label{fig:rdma_flow}
\end{figure}

With RDMA, transfers simply involve defining the source and, depending on the specific command, the destination address and the size to be transferred. Additionally, multiple transfers can be chained via the \textit{next} field of the transfer definition. These are then handled internally by the driver and the Host Channel Adapter (HCA, RDMA terminology for the network interface) and do not require further involvement from the kernel or the application.

\textit{PoCL-R} makes use of this chaining feature by having a dedicated fixed-size memory region for the command struct on both the sending side. As RDMA is currently only used for buffer transfers, the sending side sets up an \texttt{RDMA\_WRITE} to transfer the buffer contents and chains an \texttt{RDMA\_SEND} for the command struct after that.

The receiving side has a fixed buffer of command structs allocated and registered as an RDMA memory region. For each command struct in the region, a \textit{receive request} is posted to the RDMA driver. Once a buffer migration is issued from the sender, the \texttt{RDMA\_WRITE} is performed silently into the destination buffer. The chained \texttt{RDMA\_SEND} consumes a receive request from the receiver, placing the command struct's contents into one of the command struct associated with that receive request and notifying the receiver thread. All that is left for this thread is to look up the command ID from the command struct and signal the associated OpenCL events to unblock the execution of dependent OpenCL commands.

Unlike integrated GPUs, discrete GPUs typically have their dedicated global memory that might not be directly addressable by the CPU. Such GPU memory cannot trivially be registered for RDMA use, and graphics APIs usually have separate functionality for uploading data from system RAM to GPU memory. 

There is increasing support for presenting a unified virtual memory space to applications. In OpenCL, such capabilities are utilized via the Shared Virtual Memory model (SVM). However, the level and maturity of SVM support varies heavily. Therefore, \textit{PoCL-R} keeps a "scratch" or "shadow" buffer that is registered for both incoming and outgoing RDMA transfers. Upon receiving a transfer completion notification, the shadow buffer's contents are then uploaded to the right OpenCL buffer internally. Similarly, on the sending side, buffer contents are copied into the shadow buffer when a migration command is received, before issuing the RDMA transfer request.

A shadow copy of each buffer is not needed if the server's native OpenCL runtime supports allocating SVM buffers. Those get associated with proper addresses in the virtual address space of the application process and can thus be registered directly as RDMA memory regions. In \textit{PoCL-R} this mode of operation can be enabled with a compile-time option and in principle it should yield noticeable improvements to performance and memory usage. For the time being, since the SVM feature is not widely supported and the benchmarks we performed in this article used the shadow buffer solution instead. 

In the future, as SVM and related technologies get wider support, it is interesting to expand the cross-node DMA to support direct transfers also between GPU memories. There are APIs specifically for registering GPU memory directly as RDMA regions, such as NVIDIA GPUDirect\textregistered{}\cite{GPUDirect} and VK\_NV\_external\_memory\_rdma\cite{vulkan-spec}, which could be utilized internally.

\section{Evaluation}
\label{section:results}

The following subsections describe the latency and scalability benchmarks used to evaluate \textit{PoCL-R} and the results measured.

\subsection{Command Overhead}

With low latency being a high priority for \textit{PoCL-R}, a synthetic benchmark was created that measures the duration of invocations of practically empty kernels. The benchmark starts a timer and dispatches a kernel that does no work at all and simply returns. Then the benchmark awaits the completion of this kernel and stops the timer. Since the kernel performs no work, this gives a good idea of how much overhead is associated with dispatching kernels in the first place. This overhead consists of network traffic and bookkeeping in the runtime itself. The results of the benchmark are compared against the network round trip time measured with the \textit{ping} utility, which is commonly used for measuring network latency.

In order to even out fluctuations caused by buffering, drivers and unrelated processes running on the devices, the kernel invocation is repeated 1000 times and the results are averaged. The client application for this benchmark was running on a desktop PC with an Intel Core i5-3570K CPU and a 100 Mbps wired Ethernet connection to the servers. Two identical servers were used in the benchmark and the results are aggregated over both of them. The hardware configuration of the servers was as follows:
\begin{itemize}
\item{\textbf{GPU:} 2x NVIDIA GeForce 2080 Ti}
\item{\textbf{CPU:} AMD Ryzen Threadripper 2990wx}
\item{\textbf{LAN:} 100 Mbit Ethernet}
\end{itemize}

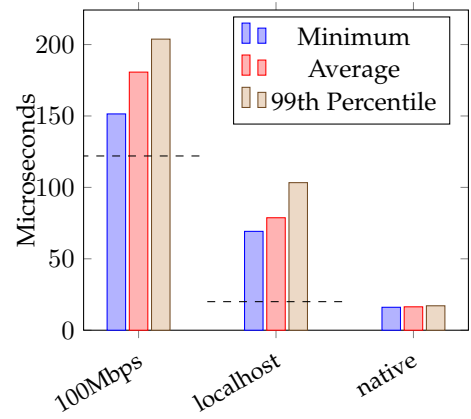
\begin{figure}
\centering
\begin{tikzpicture}[scale=0.7]
\pgfplotstableread[col sep=comma,]{command_overhead_1k_host.csv}\datatable
\begin{axis} [
    legend style={at={(1.30, 1.40)},anchor=north east},
    set layers,
    ybar,
    width=\linewidth,
    x tick label style={font=\small, rotate=30, anchor=north east},
    scaled x ticks=false,
    enlarge x limits=0.2,
    xtick=data,
    symbolic x coords={100Mbps, localhost, native},
    ylabel={Microseconds},
    ymin=0,
]
\addplot table [x=name, y expr=\thisrow{min}*1000000] {\datatable};
\addlegendentry{Minimum}
\addplot table [x=name, y expr=\thisrow{avg}*1000000] {\datatable};
\addlegendentry{Average}
\addplot table [x=name, y expr=\thisrow{p99}*1000000] {\datatable};
\addlegendentry{99th Percentile}
\addplot[black,dashed,sharp plot,update limits=false,] coordinates { ([normalized]-1,122) ([normalized]0.5,122) };
\addplot[black,dashed,sharp plot,update limits=false,] coordinates { ([normalized]0.5,20) ([normalized]1.5,20) };
\end{axis}
\end{tikzpicture}
\caption{Duration of a no-op command as measured by the client application using standard CPU timers. The dashed line represents the average ICMP ping for the scenarios that use TCP. Native refers to the native OpenCL implementation for the GPUs, provided by NVIDIA. }
\label{fig:nop}
\end{figure}

Fig.~\ref{fig:nop} shows the results of this benchmark. For reference, the ICMP round-trip latency is reported by the \textit{ping} utility to fluctuate around 0.122 ms. On the loopback interface of the PC running the client application, the ICMP round-trip was reported as 0.020 ms. OpenCL commands consistently took around 60 microseconds more than this ping latency. Since connection latency between commonplace consumer devices and application servers generally measures in tens to hundreds of milliseconds even in real-time applications, we consider 60 microseconds on top of network latency to be a good result for \textit{PoCL-R}. The overhead remains constant even when running the client application on the same machine as the server daemon, indicating that the 60 microseconds is indeed only overhead from the runtime.

Out of the identified related work, the closest that we could successfully run and compare against \textit{PoCL-R} was the latest version of SnuCL~\cite{snucl_base} (1.3.3). This framework focuses mainly on data center use and total throughput scalability, although it follows similar principles as \textit{PoCL-R}. It relies on MPI for its communications, which imposes some overhead of its own.

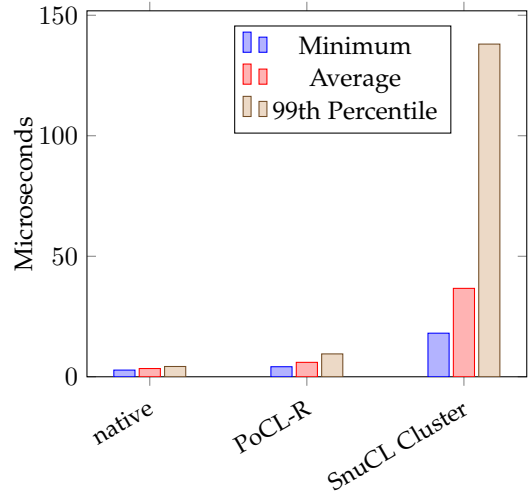
\begin{figure}
\centering
\begin{tikzpicture}[scale=0.8]
\pgfplotstableread[col sep=comma,]{passthrough_comparison.csv}\datatable
\begin{axis} [
    legend style={at={(1.00, 1.20)},anchor=north east},
    set layers,
    ybar,
    width=\linewidth,
    x tick label style={font=\small, rotate=30, anchor=north east},
    scaled x ticks=false,
    enlarge x limits=0.2,
    xtick=data,
    symbolic x coords={native, PoCL-R, SnuCL Cluster},
    ylabel={Microseconds},
    ymin=0,
]
\addplot table [x=name, y expr=\thisrow{min}*1000000] {\datatable};
\addlegendentry{Minimum}
\addplot table [x=name, y expr=\thisrow{avg}*1000000] {\datatable};
\addlegendentry{Average}
\addplot table [x=name, y expr=\thisrow{p99}*1000000] {\datatable};
\addlegendentry{99th Percentile}
\end{axis}
\end{tikzpicture}
\caption{Comparison of the runtime duration of a simple pass-through kernel on the native NVIDIA OpenCL driver,\textit{PoCL-R}and SnuCL as reported by the OpenCL event profiling API}
\label{fig:vs_snucl}
\end{figure}

In order to compare the minimum runtime latencies of SnuCL and \textit{PoCL-R}, a simple pass-through kernel was benchmarked. This kernel copies a single integer from an input buffer to an output buffer. The run times of the kernel as reported by the OpenCL event profiling API were measured for three different setups: running on the proprietary NVIDIA driver without any distribution layer, using the SnuCL Cluster runtime, and using \textit{PoCL-R}. The differences between the distributed runtimes stem from internal command management on top of the native OpenCL driver and, in the case of SnuCL, from the communication overhead from the MPI runtime. The results, as seen in Fig.~\ref{fig:vs_snucl}, place \textit{PoCL-R} well ahead with commands only taking $\frac{1}{6}$ of the time they took with SnuCL. However, when compared to calling the NVIDIA driver directly, \textit{PoCL-R} still takes twice as long, so there is certainly still room for improvement.

\subsection{Data Migration Overhead}

The authors of SnuCL report the bottleneck in some of their benchmarks to be data movement~\cite{snucl_base}. It is thus interesting to measure what the minimum duration for a buffer migration is with the two examined frameworks. \textit{PoCL-R} supports migrating buffers directly between servers in a P2P fashion, where the client application only sends migration commands to the source server of the migration. The server then pushes the data directly to the destination, which notifies the rest of the setup of the command completion. As such, this benchmark was conducted separately from the no-op command overhead measurements.

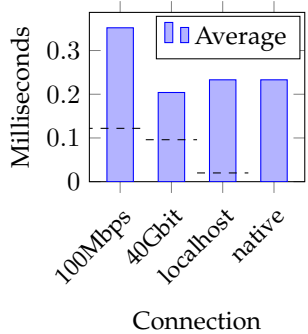
\begin{figure}[t]
\centering
\begin{tikzpicture}
\pgfplotstableread[col sep=comma,]{migration_overhead.csv}\datatable
\begin{axis} [
    set layers,
    ybar,
    width=0.5\linewidth,
    x tick label style={font=\small, rotate=45, anchor=north east},
    scaled x ticks=false,
    symbolic x coords = {100Mbps, 40Gbit, localhost, native},
    xtick=data,
    enlarge x limits=0.2,
    ylabel={Milliseconds},
    ymin=0,
    xlabel={Connection},
]
\addplot table [x=name, y expr=\thisrow{avg}*1000] {\datatable};
\addlegendentry{Average}
\addplot[black,dashed,sharp plot,update limits=false,] coordinates { ([normalized]-1,0.122) ([normalized]0.5,0.122) };
\addplot[black,dashed,sharp plot,update limits=false,] coordinates { ([normalized]0.5,0.096) ([normalized]1.5,0.096) };
\addplot[black,dashed,sharp plot,update limits=false,] coordinates { ([normalized]1.5,0.02) ([normalized]2.5,0.02) };
\end{axis}
\end{tikzpicture}
\caption{Duration of a migration of a 4-byte buffer between two
devices using different connectivity between servers, as well as using the native NVIDIA driver for reference. Numbers are averaged across 1000 migrations. The dashed line represents the average ICMP ping for the given connection.}
\label{fig:migration}
\end{figure}

The benchmark enqueues 1000 migrations between the servers and computes the average from the measured durations. The buffer being migrated only contains a single 4-byte integer in order to minimize the effect of the network speed on the results. A simple kernel that increments the integer value is enqueued between migrations to ensure that the migration really has to take place. The results are shown in Fig.~\ref{fig:migration}. With a 100 Mbps wired Ethernet connection between the servers, the timings average to roughly 3 times the overhead of a no-op command plus network ping. This seems reasonable for a 3-step round trip (client to the first server, first to the second server, second server to client) and a bit of buffer management on the servers.

Using a 40 Gbps direct link between the servers instead of a 100 Mbps link with a switch reduces the total time considerably. This is mostly because of the dedicated direct link without any other network equipment or interference from other traffic. The measurements were also performed with two \textit{PoCL-R} daemons running on the same machine and with one daemon migrating buffers between two GPUs on the same machine. The native OpenCL implementation used by the daemon turned out to suffer a noticeable performance regression when using two GPUs simultaneously, so the latter configuration did not end up being a useful comparison target.

We also attempted a comparison against SnuCL, but calls to \textit{clEnqueueMigrateMemObjects} consistently caused SnuCL to segfault, so a detailed comparison was not possible.

Two identical machines were used for benchmarking, with the following hardware configuration:
\begin{itemize}
\item{\textbf{GPU:} 2x Geforce 2080 Ti}
\item{\textbf{CPU:} Ryzen Threadripper 2990wx}
\item{\textbf{LAN:} 100Mbit Ethernet, 40Gbit direct Ethernet link between machines}
\end{itemize}

\subsection{RDMA vs. TCP Socket Communication}

The performance of RDMA for buffer migration between servers was measured with a synthetic benchmark that waits for all prior OpenCL commands (on other devices) to finish executing, starts a timer, issues an explicit migration command to get the buffer to this device and waits again until it has finished and stops the timer. Finally, the application dispatches a kernel that increments the first (integer) element in the buffer in order to invalidate other copies of the buffer before it gets migrated to the next device. This sequence loops through all devices in the OpenCL context N times.

\begin{figure}
\centering
\begin{tikzpicture}[scale=0.8]
\pgfplotstableread[col sep=comma,]{rdma_migration_speedup.csv}\datatable
\begin{axis} [
    legend style={at={(0.05,1.2)},anchor=north west}, 
    set layers,
    width=\linewidth,
    x tick label style={font=\small, rotate=30, anchor=north east},
    xmode=log,
    log basis x=2,
    xmin=2^0, xmax=2^35,
    enlarge x limits=false,
    max space between ticks = 20,
    ymajorgrids=true,
    ylabel={\% Speedup with RDMA},
    xlabel={buffer size (bytes)},
]
\addplot table [x expr=\thisrow{elements}*4, y expr=\thisrow{avg} * 100] {\datatable};
\addlegendentry{Average}
\addplot table [x expr=\thisrow{elements}*4, y expr=\thisrow{p99} * 100] {\datatable};
\addlegendentry{99th Percentile}
\end{axis}
\end{tikzpicture}
\caption{Speedup gained from using RDMA when migrating a uint32 buffer between two servers. Numbers in each configuration are aggregated over 200 separate migrations. Timings are measured in the main application, starting from clEnqueueMigrateMemObjects until clFinish indicates that the command is completed.}
\label{fig:migration-speedup}
\end{figure}
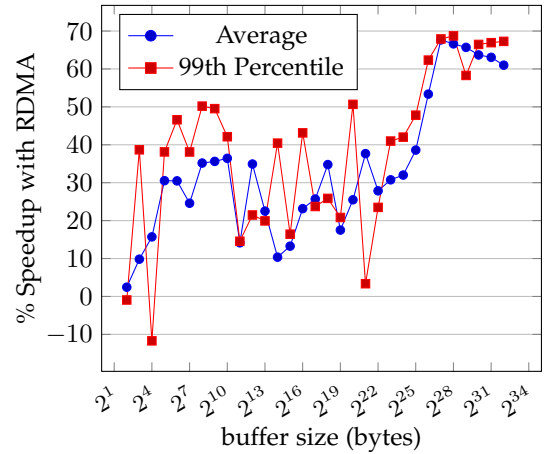

Figure \ref{fig:migration-speedup} displays the relative speedup gained from using RDMA versus the old TCP socket communication scheme when migrating a buffer between two servers with one GPU each. The timings are aggregated across 100 full migration cycles times, yielding a total of 200 migrations for each buffer size.
Since the kernel only increments the first element, the buffer size has no effect on anything except transfer duration.

RDMA becomes almost 30\% faster than the naive TCP socket solution already by the time the buffer size reaches 32 bytes. There is a fair amount of noise, presumably related to other traffic on the client-server network. As the size of the buffer being transmitted exceeds 9 MiB, the relative speedup begins to increase. This 9 MiB is also the internal send buffer size configured on the TCP socket, which suggests that this is the point where writes to the socket are getting split up and the overhead of making more and more write calls becomes apparent. The relative improvement of RDMA continues until it plateaus out at around 65\% for 134 MiB and larger buffers.

\subsection{Distributed Large Matrix Multiplication}

Scalability under non-trivial workloads was measured with a distributed matrix multiplication test. This benchmark multiplies two $N \times N$ matrices using all the available devices in the OpenCL context. The full input data is uploaded to each device, and each device computes a roughly equal number of rows of the result matrix. The actual computation is embarrassingly parallel, but the intermediate results from each device have to be collected into a single buffer to form the final result, which makes scaling this workload as a whole non-trivial.

The benchmark is broadly the same as the matrix multiplication used by SnuCL authors~\cite{snucl_base}, except that combining the partial results into a final output matrix is included in the host timings here. The example code by NVIDIA that \cite{snucl_base} mentions as the source for their benchmark only measures the duration of the multiplication kernels themselves. Whether the SnuCL benchmark accounts for combining the intermediate results is unknown, but considering the reported scalability problems, it stands to assume that combining the results was indeed part of what was measured.

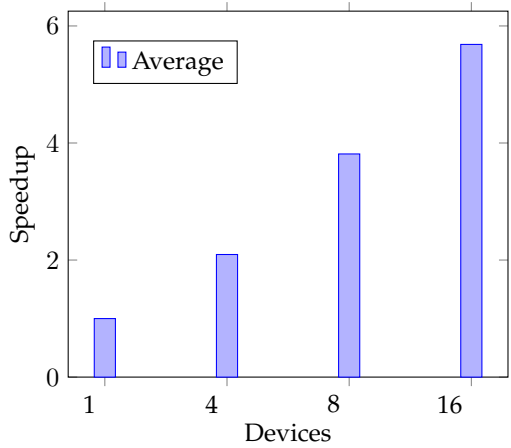
\begin{figure}
\centering
\begin{tikzpicture}[scale=0.8]
\pgfplotstableread[col sep=comma,]{matmul_cluster_8192.csv}\datatable
\begin{axis} [
    legend style={at={(0.05,1.15)},anchor=north west}, 
    set layers,
    ybar,
    width=\linewidth,
    x tick label style={font=\normalsize, rotate=0, anchor=north east},
    scaled x ticks=false,
    xtick=data,
    symbolic x coords ={1, 4, 8, 16},
    ylabel={Speedup},
    ymin=0,
    xlabel={Devices},
]
\addplot table [x=devices, y expr=17.5777/\thisrow{avg}] {\datatable};
\addlegendentry{Average}
\end{axis}
\end{tikzpicture}
\caption{Multiplication of two $8192 \times 8192$ matrices using 1 to 16 remote devices in servers with 4 GPUs each, averaged across 5 runs that were executed in parallel. Displayed is the speedup compared to using a single GPU.}
\label{fig:matmul}
\end{figure}

The benchmarks were run on a compute cluster consisting of three servers with an \textit{Intel\texttrademark{} Xeon\texttrademark{} E5-2640 v4} CPU and four \textit{NVIDIA Tesla P100} GPUs. The number of GPUs was padded to a total of 16 by adding one more server, with an \textit{Intel\texttrademark{} Xeon\texttrademark{} Silver 4214} CPU and four \textit{NVIDIA Tesla V100} GPUs. All servers, as well as the machine running the client application, were connected to a 56 Gbps LAN.

The relative speedup compared to performing the entire computation on a single GPU is shown in Fig.~\ref{fig:matmul}. The performance increase appears to follow a logarithmic curve as more GPUs are added, ending up at slightly less than 6x when using all 16 GPUs. This is comparable to the results reported by~\cite{snucl_base} when the proposed MPI collective communication extensions are used. \textit{PoCL-R} does however not exhibit the performance regression when using more than 8 devices that in SnuCL authors report.

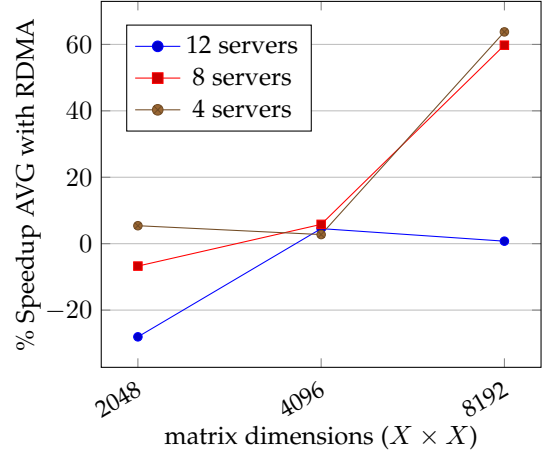
\begin{figure}
\centering
\begin{tikzpicture}[scale=0.8]
\begin{axis} [
    legend style={at={(0.05,1.15)},anchor=north west},
    set layers,
    width=\linewidth,
    xtick=data,
    x tick label style={font=\small, rotate=30, anchor=north east, /pgf/number format/precision = 4},
    symbolic x coords = {2048, 4096, 8192},
    ymajorgrids=true,
    ylabel={\% Speedup AVG with RDMA},
    xlabel={matrix dimensions ($X \times X$)},
]
\addplot table [x=elements, y expr=\thisrow{avg} * 100, col sep=comma] {
avg,min,max,p99,elements,devices
-0.28051543509804183,-0.13279738018226,-0.4284360908945783,-0.4284360908945783,2048,12
0.04532978147348239,0.10928965156012274,-0.023139295204509983,-0.023139295204509983,4096,12
0.007571925873379946,0.016540195708329226,-0.0021098858223094247,-0.0021098858223094247,8192,12
};
\addlegendentry{12 servers}
\addplot table [x=elements, y expr=\thisrow{avg} * 100, col sep=comma] {
avg,min,max,p99,elements,devices
-0.06746151710522266,0.14983874703775635,-0.010980253932739234,-0.010980253932739234,2048,8
0.05817259030747411,0.13177644400919197,-0.025482789806883923,-0.025482789806883923,4096,8
0.5976869163268758,0.6748274177180299,0.5216993243031681,0.5216993243031681,8192,8
};
\addlegendentry{8 servers}
\addplot table [x=elements, y expr=\thisrow{avg} * 100, col sep=comma] {
avg,min,max,p99,elements,devices
0.05411111185894632,0.13407753196671404,-0.023716488054972056,-0.023716488054972056,2048,4
0.02745657328875686,0.053957317892525135,-0.0029003127357420965,-0.0029003127357420965,4096,4
0.6377210352176746,0.840113209566775,0.4426161904881827,0.4426161904881827,8192,4
};
\addlegendentry{4 servers}
\end{axis}
\end{tikzpicture}
\caption{Average speedup gained from using RDMA for distributed matrix multiplication using N servers.}
\label{fig:matmul-cluster-speedup}
\end{figure}

Fig.~\ref{fig:matmul-cluster-speedup} shows the benchmark results for naive distributed matrix multiplication. Since uploading the inputs is not part of the timing calculations, only the multiplications themselves and merging the results are considered for this benchmark. The amount computed and transferred is divided equally among all servers. For small work sizes RDMA provides no meaningful speedup here and with a large number of servers the registration of RDMA memory regions and informing peers of the keys for said regions the result is even a net negative. With matrix sizes below $8192 \times 8192$, the resulting buffer size does not exceed the roughly 23 MB that Fig.~\ref{fig:migration-speedup} indicates as the tipping point where RDMA consistently reaches much better performance. In the $8192 \times 8192$ case, using 12 servers brings the buffer handled by an individual server just below this limit, hence no meaningful improvement is seen there. When only 8 or 4 servers are used, a roughly 60\% performance improvement is seen, similar to the result suggested by the synthetic buffer migration benchmark.

\section{Real-World Application Case Studies}
\label{sec:case-studies}

The following subsections describe the case studies performed with \textit{PoCL-R} in real-world applications.
The first case study is an augmented reality application where rendering quality was enhanced by optionally offloading computation to a MEC server using \textit{PoCL-R}. The second case demonstrates the scalability of the runtime at server-side in order to support applications with high performance requirements.

\subsection{Real-time Point Cloud Augmented Reality Rendering}
\label{sec:ar-rendering}

\begin{figure}
    \centering
    \includegraphics[width=0.4\linewidth]{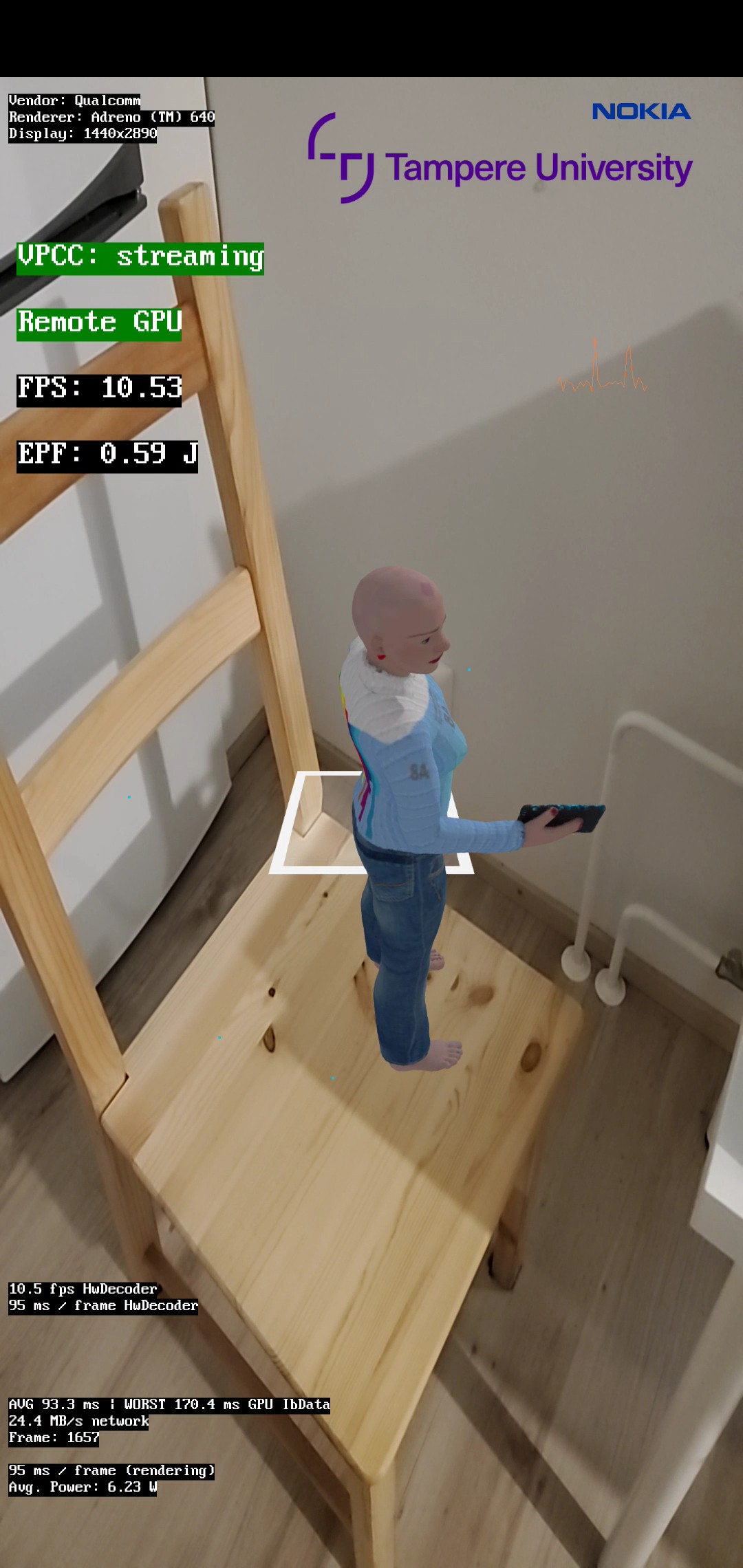}
    \caption{Screenshot of the AR application used to measure the effect of offloading heavy computation. A streamed animated point cloud of a person holding a small tablet device is displayed in augmented reality on top of a real-world chair.}
    \label{fig:ARdemo_Screen}
\end{figure}

For this case study, we implemented an Android-based smartphone application with real-time AR rendering.
A sample screenshot of the application is shown in Fig.\ref{fig:ARdemo_Screen}. 

The point cloud is received as a Video-based Point Cloud Compression (VPCC) stream encoded with HEVC~\cite{h265,hevc-overview}. The video data is decompressed with a hardware HEVC decoder present in the smartphone SoC. The point cloud is then reconstructed from the data with OpenGL~\cite{gles-spec} shaders~\cite{glsl-es-spec}. A more thorough explanation of the reconstruction process is given in \cite{nokia_ar}.

The points of the received point cloud are sorted by their distance from the viewer in order to enable the use of alpha blending when rendering the points. Alpha blending produces less pixelated results than simply drawing the points as squares, yielding a much more visually pleasing end result.

Compared to reconstructing the individual point positions, sorting the points by distance to an arbitrary point is a very computationally heavy task, so it is a prime candidate for offloading to a high-powered compute server. When offloading is not in use, the steps from decoding the VPCC stream, to reconstructing and reordering the points, to the final render are all ran on the smartphone SoC. When offloading is enabled, however, the compressed VPCC stream is sent to both the smartphone and the compute server. Both devices decode and reconstruct the points, but only the remote server computes the sorted order of the points. The server then sends a sorted list of point indices to the smartphone to be used for rendering the point cloud. This way, the mobile SoC is free to perform other tasks such as pose estimation for tracking the position of its own viewpoint based on the live image from the smartphone's built-in camera.

OpenCL classifies compute devices as CPU, GPU, or Accelerator. With OpenCL 1.2, a new type was added, called \texttt{CL\_DEVICE\_TYPE\_CUSTOM}. This allows an implementation to expose arbitrary functionality as an OpenCL device that does not support building OpenCL C kernels provided by the application, but only allows dispatching a number of built-in kernels\cite{opencl-api-spec}. The \textit{PoCL-R} daemon exposes the video decoding hardware in the GPU of the server as such a device. To the application, this appears as just another OpenCL device that supports a "decode" kernel for decoding HEVC data from a buffer to raw pixels in another buffer. Once decoding is done and the resulting pixel buffer is usable, an OpenCL command completion event is generated to allow further OpenCL commands to use the decoded data. Since the event-based scheduling happens entirely on the server, there is no need to wait for the network round-trip to the UE. The dynamic buffer size extension proposed in \cite{pocl2021samos} can additionally be used to massively reduce the bandwidth needed to transfer the compressed HEVC buffers between devices. Another improvement is to implement another custom device that writes VPCC stream directly from its source to an OpenCL buffer that can be moved around with the optimized buffer migration mechanism. In this case study, the source is a prerecorded file. The custom streaming device provides a built-in kernel that can be accessed via \texttt{clCreateProgramWithBuiltInKernels} that writes the next chunk of the stream to an application-defined OpenCL buffer.

\begin{figure}
\centering
\begin{tikzpicture}[scale=0.7]
    \begin{axis}[
        xtick=data,
        xticklabels from table={\resultsVpccFPS}{benchname},
        xticklabel style={font=\small, rotate=30, anchor=north east},
        axis y line*=left,
        ylabel={Frames / second},
            legend style={at={(0.05,0.95)},
                anchor=north west,legend columns=1},
            enlarge y limits={upper,value=0.2},
        ybar,
        ymin=0,
        width=\linewidth,
        bar width=9pt,
        bar shift=-6pt,
        nodes near coords,
        every node near coord/.append style={rotate=90, anchor=west}
        ]
      \addplot table [x=position,y=FPS] {\resultsVpccFPS};
      \label{ardemo_fps}
    \end{axis}
    \begin{axis}[
        xtick=data,
        xticklabels from table={\resultsVpccEPF}{benchname},
        xticklabel style={font=\small, rotate=30, anchor=north east},
        axis y line*=right,
        ylabel={Joules / Frame},
            legend style={at={(0.02,0.98)},
                anchor=north west,legend columns=1},
            enlarge y limits={upper,value=0.2},
        ybar,
        ymin=0,
        width=\linewidth,
        cycle list shift=1,
        bar width=9pt,
        bar shift=6pt,
        nodes near coords,
        every node near coord/.append style={rotate=90, anchor=west}
        ]
      \addlegendimage{/pgfplots/refstyle=ardemo_fps}\addlegendentry{FPS}
      \addplot table [x=position,y=EPF] {\resultsVpccEPF};
      \addlegendentry{EPF}
    \end{axis}
\end{tikzpicture}

\caption{Frame rate and energy consumption (normalized to frame rate, Energy Per Frame) of the AR demo application in various offloading configurations. The lGPU and rGPU identifiers refer to the mobile device's local GPU and the remote GPU exposed via \textit{PoCL-R} respectively. AR indicates that AR position tracking is being used. P2P refers to the peer-to-peer buffer migration feature in \textit{PoCL-R} being used, and DYN indicates that the buffer content size extension is used to reduce network traffic between remote devices.}
\label{fig:ARdemo_perf}
\end{figure}
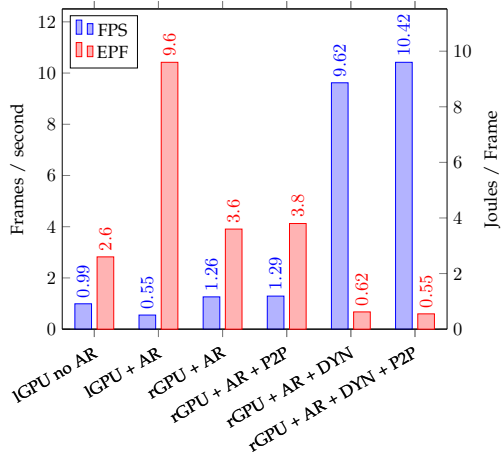

Fig.~\ref{fig:ARdemo_perf} displays the frame rate achieved by the demo application in various configurations. The first two bars represent performance when using only the GPU integrated in the device's SoC for reconstructing the points of the point cloud, sorting them and computing the camera position for AR rendering. For the next two numbers, sorting of the points was performed on a \textit{PoCL-R} remote server, both with buffer migrations performed in P2P fashion and with a host round-trip. Compared to performing all operations on the mobile UE, this already yields a 2.3x speedup. Finally, the effect dynamic buffer size extension is introduced to the application, drastically reducing the amount of data that needs to be transferred over the network and improving the frame rate almost 19x compared to performing all work on the UE.

The energy consumption on the smartphone was measured using the Android Power Stats HAL interface. The results are shown in
Fig.~\ref{fig:ARdemo_perf}. Since simply offloading the sorting step to a server almost entirely offsets the huge spike in power consumption caused by adding AR tracking, it suggests that the SoC was switching itself to a high power state to compensate for the increased load, while offloading the sorting step allowed it to stay at a lower power state while maintaining similar performance. Peer-to-peer buffer migrations appear to have a minor impact here, but the buffer content size extension helps bring energy consumption on the smartphone down to only around 20\% of that of sorting the points locally and rendering them without AR tracking.

While the compute performance of the local GPU is sufficient for tracking AR camera movement and rendering static objects, when doing the same with animated objects of reasonable detail the frame rate is lackluster. Here, \textit{PoCL-R} serves as an enabler technology, bringing the frame rate into a much more usable range and reducing power consumption per frame by a factor of 2.5x in the worst case to 17x in the optimized case. The reduced energy consumption per frame creates the possibility of deliberately reducing the frame rate in order to improve battery life on handheld devices.

Hardware used in this test:
\begin{itemize}
\item{\textbf{Remote GPU:} GeForce 1060 3 GB, Intel Core i7-6700}
\item{\textbf{Remote custom device:} a virtual device implemented as a \textit{PoCL} device driver simulating a point cloud camera by reading the stream from a file. Acts as the data source for the application.}
\item{\textbf{Mobile device:} Samsung Galaxy S10 SM-G973U1, Qualcomm\textregistered{} Snapdragon\texttrademark{} 855, connected via Wi-Fi 6}
\item{\textbf{Wi-Fi router:} ASUS ROG Rapture GT-AX11000}
\item{\textbf{Wi-Fi router to remote server connection:} 1Gbit wired Ethernet}
\end{itemize}

\pgfplotstableread{
benchname                                position        FPS
{\shortstack{lGPU no AR}}                  1             0.99
{\shortstack{lGPU + AR}}                   2             0.55
{\shortstack{rGPU + AR}}                   3             1.26
{\shortstack{rGPU + AR + P2P}}             4             1.29
{\shortstack{rGPU + AR + DYN}}             5             9.62
{\shortstack{rGPU + AR + DYN + P2P}}       6             10.42
}\resultsVpccFPS

\pgfplotstableread{
benchname                                position        EPF
{\shortstack{lGPU no AR}}                  1             2.6
{\shortstack{lGPU + AR}}                   2             9.6
{\shortstack{rGPU + AR}}                   3             3.6
{\shortstack{rGPU + AR + P2P}}             4             3.8
{\shortstack{rGPU + AR + DYN}}             5             0.62
{\shortstack{rGPU + AR + DYN + P2P}}       6             0.55
}\resultsVpccEPF

\subsection{Multi-Node Computational Fluid Dynamics}
\label{sec:fluidx3d}

Server side multi-node scalability in real-world scenarios was measured using the FluidX3D\cite{Lehmann_FluidX3D_2022} tool. FluidX3D simulates dynamic fluid behavior using the lattice Boltzmann method and a number of optional extensions for various specialized scenarios.\cite{Lehmann_phd}.

High-Performance Computing (HPC) applications are generally run in large compute clusters, where multi-node distribution is implemented using an MPI. Also, FluidX3D has been scaled to multi-node using MPI in the past~\cite{fluidx3d-mpi}. Using MPI, however, might require  changes to the application code and is generally not available outside HPC clusters, which causes a bit of a split between HPC and non-HPC applications. In contrast, \textit{PoCL-R} works with any OpenCL application without changes to application code. However, running applications with \textit{PoCL-R} can greatly magnify inefficiencies stemming from certain inadvisable (although still valid) API usage patterns, such as the following:

FluidX3D supports splitting up the simulation into multiple domains in order to spread computation across multiple compute devices. However, the boundaries of domains have to be synchronized across devices after each simulation time step. The basic implementation does this by manually downloading the boundary region from each device into RAM and uploading it to the destination device. This allows it to work across devices that are not part of the same OpenCL platform (e.g., because they are from different vendors) but is not ``idiomatic OpenCL'' and prevents OpenCL implementations from optimizing data transfers. To fix this, a new mode was added where the buffers that hold the boundary region data are implicitly migrated to the destination device, allowing the OpenCL implementation to perform the transfers in the most optimal way it is capable of. For \textit{PoCL-R}, this means using the peer-to-peer connections between servers or, if both devices reside on the same server, deferring the task to the server's native OpenCL driver.

The test setup for the server-side scalability benchmark was as follows:
\begin{itemize}
    \item \textbf{Local device:} 1x desktop PC with a gigabit network interface
    \item \textbf{Remote servers:}: 3x servers with 100 Gbps fiber optic interfaces and NVIDIA RTX A6000 GPUs
    \item \textbf{Router:} A 100 Gbps switch that connects all 4 machines
\end{itemize}
FluidX3D itself was run in benchmark mode which runs only the basic lattice Boltzmann simulation without extensions and visualization. For the TCP and RDMA benchmarks, the application was run on the desktop PC. For benchmarks on localhost (i.e., client and server on the same machine) and native (using the NVIDIA OpenCL driver directly, without \textit{PoCL-R}) the application was run on one of the servers.

\begin{figure}
\centering
\begin{tikzpicture}[scale=0.8]
\begin{axis} [
    legend style={at={(0.05,1.15)},anchor=north west},
    set layers,
    width=\linewidth,
    xtick=data,
    ymajorgrids=true,
    ylabel={Peak MLUPs/s},
    xlabel={Number of GPUs},
]
\addplot table [x=ngpus, y expr=\thisrow{mlups}, col sep=comma] {
setup, ngpus, grid,   mlups, util
  tcp,     3,  514,    9017,   85
  tcp,     2,  514,    5897,   78
  tcp,     1,  514,    3750,   98
};
\addlegendentry{TCP only}
\addplot table [x=ngpus, y expr=\thisrow{mlups}, col sep=comma] {
setup, ngpus, grid,   mlups, util
 rdma,     3,  514,    9096,   82
 rdma,     2,  514,    6211,   81
  tcp,     1,  514,    3750,   98
}; 
\addlegendentry{RDMA}
\addplot table [x=ngpus, y expr=\thisrow{mlups}, col sep=comma] {
setup, ngpus, grid,   mlups, util
local,     3,  514,    8572,   78
local,     2,  514,    6235,   85
local,     1,  514,    3808,  100
};
\addlegendentry{Localhost}
\addplot table [x=ngpus, y expr=\thisrow{mlups}, col sep=comma] {
setup, ngpus, grid,   mlups, util
   nv,     3,  514,    8905,   83
   nv,     2,  514,    6208,   87
   nv,     1,  514,    3822,  100
};
\addlegendentry{NVIDIA}
\end{axis}
\end{tikzpicture}
\caption{FluidX3D performance, in Millions of lattice updates per second. NVIDIA represents using the vendor driver directly (all GPUs in the same machine).}
\label{fig:fluidx3d-performance}
\end{figure}
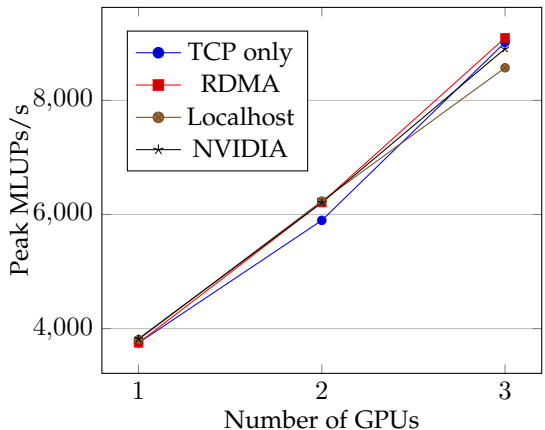

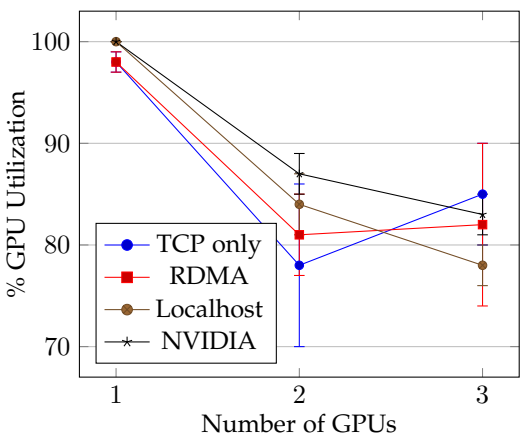
\begin{figure}
\centering
\begin{tikzpicture}[scale=0.8]
\begin{axis} [
    legend style={at={(0.025,0.025)},anchor=south west},
    set layers,
    width=\linewidth,
    xtick=data,
    ymajorgrids=true,
    ylabel={\% GPU Utilization},
    xlabel={Number of GPUs},
]
\addplot+ [error bars/.cd, y dir=both, y explicit] table [x=ngpus, y expr=\thisrow{util}, y error=util_err, col sep=comma] {
setup, ngpus, grid,   mlups, util, util_err
  tcp,     3,  514,    9017,   85,        5
  tcp,     2,  514,    5897,   78,        8
  tcp,     1,  514,    3750,   98,        1
};
\addlegendentry{TCP only}
\addplot+ [error bars/.cd, y dir=both, y explicit] table [x=ngpus, y expr=\thisrow{util}, y error=util_err, col sep=comma] {
setup, ngpus, grid,   mlups, util, util_err
 rdma,     3,  514,    9096,   82,        8
 rdma,     2,  514,    6211,   81,        4
  tcp,     1,  514,    3750,   98,        1
};
\addlegendentry{RDMA}
\addplot+ [error bars/.cd, y dir=both, y explicit] table [x=ngpus, y expr=\thisrow{util}, y error=util_err, col sep=comma] {
setup, ngpus, grid,   mlups, util, util_err
local,     3,  514,    8572,   78,        2
local,     2,  514,    6235,   84,        3
local,     1,  514,    3808,  100,        0
};
\addlegendentry{Localhost}
\addplot+ [error bars/.cd, y dir=both, y explicit] table [x=ngpus, y expr=\thisrow{util}, y error=util_err, col sep=comma] {
setup, ngpus, grid,   mlups, util, util_err
   nv,     3,  514,    8905,   83,        2 
   nv,     2,  514,    6208,   87,        2
   nv,     1,  514,    3822,  100,        0
};
\addlegendentry{NVIDIA}
\end{axis}
\end{tikzpicture}
\caption{FluidX3D GPU utilization, with 1 GPU per server node. NVIDIA represents using the vendor driver directly and Localhost running server and client on the same machine. In these two cases, all GPUs are on the same server node.}
\label{fig:fluidx3d-gpu-util}
\end{figure}

Fig.~\ref{fig:fluidx3d-performance} shows that performance with \textit{PoCL-R} scales with the number of server nodes almost as well as the NVIDIA driver scales with the number of local GPUs. In terms of GPU utilization, there is a slight overhead from using \textit{PoCL-R} as seen in Fig.~\ref{fig:fluidx3d-gpu-util}, but in comparison to the overhead of using multiple GPUs on the same machine, it holds up relatively well. The 'Localhost' line shows that the majority of this overhead clearly comes from the network speed, as running the simulation through \textit{PoCL-R} with the server and client on the same machine yields throughput that is within the usual fluctuation of what is achieved when using the NVIDIA driver directly.

As it turns out, RDMA does not benefit this benchmark much, since even with the largest grid that the NVIDIA driver allows allocating in one chunk without memory compression\cite{Lehmann2022_lbm_float_formats} ($514 \times 514 \times 514$ per used GPU), the boundary buffers are only around 5.2 MB in size. This means that the entire buffer fits in the kernel-side send and receive buffers of the TCP stack (configured to 9 MB for buffer transfers) and can be sent or received with a single system call. Some improvement might be visible by eliminating the CPU altogether by performing RDMA transfers directly from GPU memory to GPU memory instead of going through the CPU. It also appears that the NVIDIA driver performs device to device copies by circulating them through the main memory instead of utilizing PCIe peer-to-peer copies, making it scale to multiple GPUs roughly as well as \textit{PoCL-R} scales to multiple server nodes as seen in Fig.~\ref{fig:fluidx3d-performance}. The lack of PCIe peer-to-peer copies also hides the aforementioned inefficiency of manually circulating buffer data through the CPU. In the future, performing server-to-server transfers with RDMA directly from GPU memory to GPU memory could further improve performance scaling.

During the initialization phase, the application is heavily bottle necked by the desktop's gigabit connection, as that was fully saturated for the majority of that phase. During the simulation itself client traffic was a steady 7 KiB/s per used GPU and the p2p traffic between servers measured at 231 MiB/s (slightly under 2 Gbps) RX/TX on each server, summing up to a combined bandwidth of around 12 Gbps passing through the switch with 3 server nodes. This highlights the importance of p2p transfers, since routing 12 Gbps of traffic through the client application when copying data from server to server is impractical at best.

As shown in Fig.~\ref{fig:fluidx3d-gpu-util}, multi-node GPU utilization is in the order of $80\%$, which matches the MLUPs/s increase seen in Fig.~\ref{fig:fluidx3d-performance}. This is comparable to the scaling results of the MPI port\cite{fluidx3d-mpi}, suggesting that \textit{PoCL-R} a viable alternative to MPI as a distribution layer for some applications, as far as performance is concerned. Additionally, \textit{PoCL-R} having light requirements on the system opens up new possible scenarios, such as running an unmodified HPC application like FluidX3D on handheld UEs while performing the actual computation somewhere with enough power to do it in real time.

\section{Conclusions}
\label{section:conclusions}

In this paper, we proposed \textit{PoCL-R}, a distributed OpenCL-based heterogeneous computing runtime 
which both minimizes response latency and optimizes multi-node scalability on the server side that can cope with connection loss.
Performance enhancing features of \textit{PoCL-R} include RDMA accelerated peer-to-peer buffer transfers between remote computers participating in the execution and server-side orchestration of multi-device execution across nodes.

The suitability of \textit{PoCL-R} for real-time MEC offloading was evaluated with an AR point cloud renderer running on a smartphone. Offloading is shown to yield up to 19x better frame rates at around 5.7\% of the energy consumption when the optimizations outlined in this article were applied.

Performance scalability to support demanding workloads was demonstrated with a computational fluid dynamics simulation, where a multi-node efficiency of around 80\% was observed. This was found to be comparable to the performance scaling of an MPI port of the same application\cite{fluidx3d-mpi}, suggesting that \textit{PoCL-R} can also remove the need for MPI in some multi-node cases.

The results show that the proposed offloading layer 
can serve as a compute layer for heterogeneous distributed edge offloading cases which have both high performance and low latency requirements. Since OpenCL is now increasingly used as a \textit{computing layer} below high-level programming models such as OpenVX\cite{OpenVX}, SYCL\cite{sycl-spec} and CUDA/HIP\cite{chipstar}, we believe the open source \textit{PoCL-R} runtime will have wide utility in the future.

\bibliographystyle{IEEEtran}
\bibliography{pocl-r}

\begin{IEEEbiography}[{\includegraphics[width=1in,height=1.25in,clip,keepaspectratio]{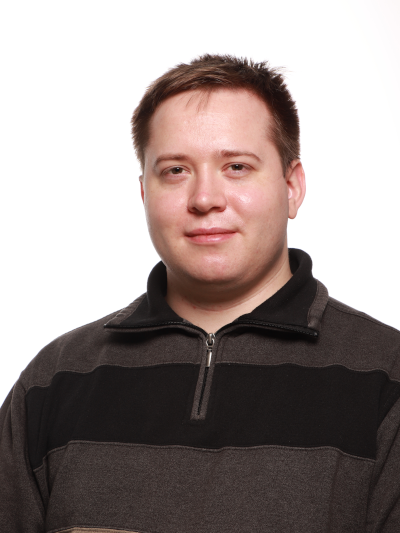}}]{Jan Solanti} has been programming since the mid 2000s. He has worked professionally on graphics and HPC programming since the mid 2010s. In his free time, he contributes to various open source projects and enjoys making the pieces to fall into place so that a complex system springs to life, like is the case with \textit{PoCL-R}. His research interests are on graphics rendering and on fast networked systems that do not depend on centralized resources.
\end{IEEEbiography}

\begin{IEEEbiography}[{\includegraphics[width=1in,height=1.25in,clip,keepaspectratio]{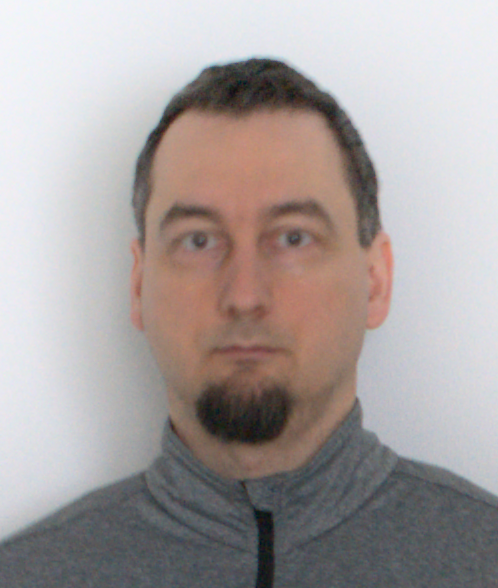}}]{Michal Babej} has been working with open-source software since 2007 and with OpenCL since 2013. He is a leading contributor to the open source PoCL project, with interests in HPC and heterogeneous computing software stacks.
\end{IEEEbiography}

\begin{IEEEbiography}[{\includegraphics[width=1in,height=1.25in,clip,keepaspectratio]{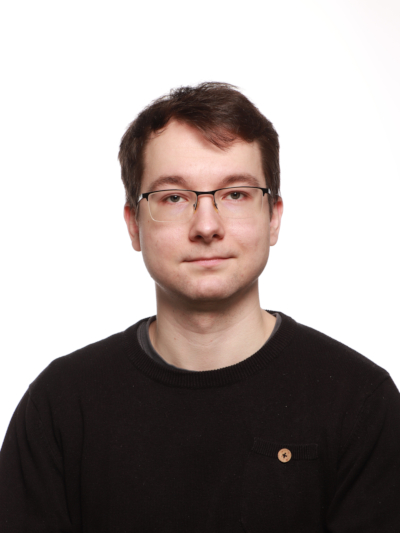}}]{Julius Ikkala} received the master’s degree in information technology from Tampere University in 2021 and is now pursuing a doctoral degree. Julius is a long-time contributor to computationally challenging open source projects whose research interests include photorealistic real-time rendering, especially ray tracing.
\end{IEEEbiography}

\begin{IEEEbiography}[{\includegraphics[width=1in,height=1.25in,clip,keepaspectratio]{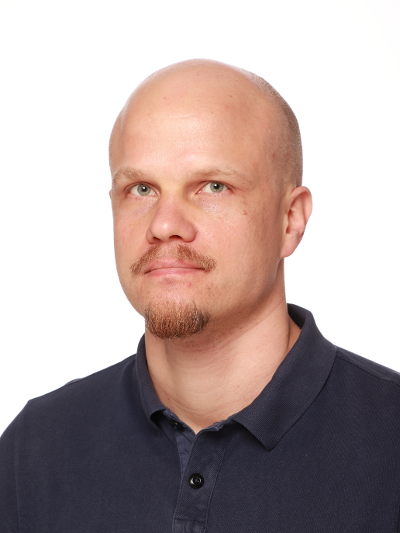}}]{Pekka J{\"a}{\"a}skel{\"a}inen} (Assoc. Professor) has worked on parallel heterogeneous platform design and programming topics since the early 2000s. In addition to his publication activities, he leads the open source development of heterogeneous computing related open source projects such as PoCL and OpenASIP. He is interested in implementation challenges of applications with both low latency and high-performance requirements, and, on the other hand, the smallest scale of computing use cases with extreme energy efficiency requirements.

\end{IEEEbiography}

\end{document}